**Using Geographically Weighted Models to Explore Temporal and Spatial Varying Impacts on Commute Trip Change Due to Covid-19**

**Saeed Saleh Namadi (**saeed@umd.edu)
Graduate Research Assistant
Department of Civil and Environmental Engineering
University of Maryland, 3105 Jeong H. Kim Engineering Building, College Park, MD 20742, United States

**Behnam Tahmasbi** (behnamt@umd.edu)
Graduate Research Assistant
Department of Civil and Environmental Engineering
University of Maryland, 3105 Jeong H. Kim Engineering Building, College Park, MD 20742, United States

**Asal Mehditabrizi** (asal97@umd.edu)
Graduate Research Assistant
Department of Civil and Environmental Engineering
University of Maryland, 3105 Jeong H. Kim Engineering Building, College Park, MD 20742, United States

**Aref Darzi (**adarzi@umd.edu)
Assistant Research Engineer
Center for Advanced Transportation Technology Laboratory
University of Maryland, Technology Ventures Building, College Park, MD 20742, United States

**Deb Niemeier** (niemeier@umd.edu)
Professor
Department of Civil and Environmental Engineering
University of Maryland, 1188 Glenn Martin Hall, College Park, MD 20742, United States

**Abstract**:
COVID-19 has deeply affected people's daily life and travel behaviors. Comprehending changes in travel behavior holds significant importance, making it imperative to investigate the influential factors of socio-demographics and socioeconomics on such behavior. This study used large-scale mobile device location data at the U.S. county level in the three states of Washington, D.C., Maryland, and Virginia (DMV area) to reveal the impacts of demographic and socioeconomic variables on commute trip change. The study investigated the impact of these variables on commuter trips over time and space. It reflected the short and long-term impact of COVID-19 on travel behavior via linear regression and geographically weighted regression models. The findings indicated that counties with a higher percentage of people using walking and biking (active mode) for commuting during the initial phase of COVID-19 experienced a greater reduction in their commute trips compared to others. Conversely, for the long-term effect of COVID-19 in November, we can see the impact of using active mode on trip change is not significant anymore and instead results showed people who were using bus and rail (public mode) for commuting decreased their trips more than others. Additionally, a positive correlation was observed between median income levels and the reduction in commute trips. On the other hand, sectors that necessitated ongoing outdoor operations during the pandemic, such as manufacturing, wholesale trade, and food services, showed a substantial negative correlation with trip change. Moreover, in the DMV area, counties with a higher proportion of Democratic voters experienced less trip reduction than others. Notably, by applying the Geographically Weighted Regression (GWR) and Multiscale Geographically Weighted Regression (MGWR) model the local spatial relationships of variables and commuting behaviors were captured. The results showed the emergence of local correlations as the pandemic evolved, suggesting a geographical impact pattern. At the onset of COVID-19, the pandemic's impact on commuting behaviors was global. However, as time passed, travel behavior became more influenced by spatial factors and started to show localized effects.

# 1. Introduction

Understanding travel behavior is crucial in transportation planning, as it enables governments and agencies to allocate resources effectively for various planning applications. Historically, travel surveys have been used to collect data on individual and household travel behavior, but they require significant preparation, labor, and costs. However, over the last two decades, advancements in technology have led to the widespread availability of mobile device location data (MDLD), offering researchers and practitioners valuable insights into human travel patterns. The emergence of COVID-19 further highlighted the importance of studying the impact of the virus on the economy and mobility, with the economic consequences being profound, leading to job losses and businesses struggling to adapt [1]. Policymakers can use insights from MDLD and other metrics to identify the sectors most affected by the virus and formulate targeted disaster policies. This data-driven approach can provide more accurate assessments of travel behavior in specific regions, enhancing transportation planning.[2]

The study aims to examine the temporal and spatial impacts of COVID-19 on commute trip changes in the DMV area (Washington, D.C., Maryland, and Virginia). Data from March-April and November of 2019 and 2020 are analyzed to understand the pandemic's short and long-term effects on travel patterns. The research investigates the emergence of local correlations, indicating a geographical impact pattern as the pandemic evolved. The study follows four main steps: collecting commuting trip data, identifying key factors for analyzing trip changes, using linear regression models to study COVID-19's short and long-term effects, and assessing spatial autocorrelation for applying the Geographically Weighted Regression (GWR) and Multiscale Geographically Weighted Regression (MGWR) model.

The research employs mobile device location data, a vast dataset that offers immense insight into people's travel behavior in the US. Unlike traditional survey data with relatively few respondents, this study's mobile device location dataset encompasses a significantly larger sample. Specifically, for November 2019 and 2020, it includes 223,662 and 419,194 identified unique device IDs, respectively. Furthermore, it contains 2,887,420 and 6,567,280 unique linked trip IDs for 2019 and 2020, respectively, demonstrating the extensive breadth and depth of the data used in this investigation.

In past research, data related to job sectors were typically gathered through surveys or Point of Interest (POI) methodologies. This investigation, however, takes a different approach. It utilizes an expansive dataset that, while lacking in sociodemographic details related to the trips, does offer insight into the association between commute trips and the distribution of 9 job sectors across each county. This analysis is facilitated through the deployment of regression models, enabling the study to elucidate how varying job sectors were influenced by the COVID-19 pandemic.

This study examines the correlation between socio-demographic factors and commute trips, focusing on the temporal and spatial aspects. To do so, we also investigated the the extent of clustering and spatial autocorrelation evolution during the pandemic. Utilizing extensive mobile device location data, the research offers a perspective on these spatial relationships, presenting valuable insights for local governments. Furthermore, conducting comprehensive investigations into these connections proves advantageous in formulating effective strategies to support those impacted by COVID-19 outbreaks. Given that specific regions remain at risk of a potential new outbreak, the study's results can be instrumental in preparedness initiatives, encompassing the creation of local relief plans for any future outbreaks.

The research approach of this study starts with an extensive review of the literature on the effects of COVID-19 on commute travel behavior by survey and mobile device location data. The key research gap is identified from the literature review. Then datasets used in this study are introduced. With the datasets introduced, a methodological framework is proposed and applied to the datasets. The results further show the COVID-19 impact on commute trips.

## 2. Literature Review

The COVID-19 pandemic, which rapidly spread globally from December 2019 [3] to May 2023, affected a staggering 766 million people and claimed 6,935,889 lives, constituting a severe global health disaster [4]. In response, non-pharmaceutical intervention strategies like travel restrictions, early virus detection, physical distancing, and stay-at-home orders emerged as crucial tools to curb the virus's spread [5] [6]. The United States implemented stay-at-home orders followed by transitioning to a phased reopening process based on government guidelines. The pandemic significantly impacted travel behavior, with both non-commute and commute trips decreasing. [7] Two main approaches, survey data collection, and mobile device location data, offer insights into trip behavior changes during COVID-19.

Travel surveys are crucial for studying human mobility patterns, capturing various data like origin, destination, mode of transport, and purpose of trips, along with sociodemographic and economic details [8]. Research on travel behavior has identified sociodemographic and economic factors as significant determinants [9]. An online survey conducted in the UK by Harrington and Hadjiconstantinou [10] revealed shifts in transit choices during COVID-19, with some switching to cycling or walking while others moved from public transport to cars due to safety concerns. Lisa Ecke [11] utilized the German Mobility Panel to provide valuable insights into travel behavior during the pandemic, using detailed socio-economic metrics. Bick et al. provide insights into how many US workers transitioned to remote work in the months following the pandemic's onset [12]. The absence of detailed information, such as job titles and sociodemographic factors, can prompt researchers to merge mobile device location data and surveys. However, this integration is not possible solely with MDLD. During the initial lockdown in April 2020, the Ministry of Infrastructure and Water Management requested employed panel members from the Nederlands Verplaatsings panel (NVP), whose GPS data had been collected, to participate in an online survey [13]. Integrating mobile device location data and surveys can enhance data, but limitations exist due to privacy and sample size constraints.

Over recent years, the use of mobile device location data for examining population travel behavior has gained significant traction. Numerous studies have incorporated this type of data, such as Chankaew et al [14], who examined freight traffic through national truck GPS data in Thailand. Call detail records (CDRs), generated by telephone exchanges or similar telecommunications equipment. Cellular phone location data have been harnessed by Colak et al.[15] to construct pairs of home and work journeys and assign trip purposes. Ahas and his team [16] evaluated a model that identifies users' home and work locations by comparing Estonian mobile data with population register data. More recently, Aslam, Cheng, and Cheshire [17] employed smart card data to detect the locations of homes and workplaces, while Yang et al. [18] used mobile data to decipher the spatial structure of urban commuting. Xu et al. [19] used mobile data in their study to understand how individuals navigate in relation to key activity locations. Kung and his team [20] scrutinized patterns in home-to-work commutes as well. Recent academic works have delved into the effects of the COVID-19 pandemic on human mobility trends. Engle, Stromme, and Zhou[21] examined how local disease prevalence and stay-at-home orders influenced individual movements outside of residences. Bick, Blandin, and Mertens [12] charted the progression of remote work practices throughout the pandemic and anticipated the future adoption of this work style using cellular location data.

Furthermore, Huang et al. [22] explored the relationship between demographic and socioeconomic factors and home-stay durations following stay-at-home orders using large-scale mobile phone location tracking data. Zhang et al.[23] designed an interactive tool utilizing anonymized mobile device location data to identify trips and generate variables, such as a social distancing index, the proportion of individuals remaining at home, the frequency of visits to work and non-work locations, out-of-town journeys, and trip distances. Meanwhile, Jiao, Bhat, and Azimian [24] investigated how the count of COVID-19 cases could predict subsequent week's travel behaviors in Houston, Texas. Lastly, Pourfalatoun and Miller [25] analyzed the impacts of COVID-19 on the usage and perception of micro mobility.

The evolution of travel behavior during the COVID-19 pandemic has been tackled through diverse models and strategies. Engle et al. [21] utilized a Regression model to estimate alterations in mobility. Furthermore, the percent change in total foot traffic in Houston was examined using an Autoregressive Distributed Lag (ADL) multiple regression model [24]. Home dwelling time was scrutinized using Pearson Correlation and Random Forest Regression [22]. Tejas Santanam et al.[26] sought to understand the impact of lockdowns related to COVID-19 in the metropolitan area of Atlanta, Georgia. They scrutinized commuter patterns in three periods: pre-pandemic, during the lockdown, and post-lockdown. The Ordinary Least Squares (OLS) approach cannot evaluate spatial effects among observations and neglects spatial autocorrelation, despite prior studies affirming the spatial correlation of COVID-19 outbreaks [27]. In recent papers,(GWR) was deployed to assess the demographic and economic disparities in COVID-19 infections in light of spatial impacts. This technique can accommodate spatial autocorrelations and evaluate varying coefficients from observations instead of estimating average coefficients in OLS models [28] [29]. Junfeng Jiao and colleagues examined the spatial-temporal patterns and economic-demographic disparities in COVID-19 infections across varying population densities in the United States [30]. This inquiry involved the application of OLS, GWR, and Random Forest (RF) methodologies. Yanwen Liu et al. [31] delved into the patterns of population migration and spatial-temporal variation related to COVID-19. To comprehend the multifaceted nature of COVID-19's spread through the lens of space and time, Xiu Wu and Jinting Zhang [32] explored the geographical and temporal heterogeneity at the county level in Texas.

Several studies have been conducted regrading COVID-19 impact on travel behavior change either by survey data or mobile device location data. However, most studies have focused on all kinds of trips (not just commute trips), and long-term effects have not been explored for commute trips. Moreover, detailed job sectors were not identified in those studies that analyze commute trips. In addition, some interesting socio and economic metrics such as the percentage of their political preferences, have not been explicitly studied. The key research gap identified from the literature review indicates that few studies focused on spatial-temporal varying impacts on COVID-19 cases. No study specifically studies the arising of spatial autocorrelation of commute trip change due to COVID-19, whether the change in trips had a global or local spatial relation, and if the state of autocorrelation evolved over time.

## 3. Data and Methods

### *3.1 Dataset*
This study's primary data is from MDLD, which was procured from several leading data providers. The methodologies used in this research were developed by the University of Maryland (UMD) project team to develop national passenger OD data [33]. This dataset carries extensive spatial and temporal information of numerous users. Details include a randomly hashed device identifier, geographical coordinates (latitude and longitude) of location points, and a timestamp reflecting when the user's location was recorded. The raw data had an array of processing steps to arrive at the final dataset, encapsulating commute trips, modes of travel, trip count, and travel distance. [34] [35] [36] [37] [38] [39]

The percentage of each job sector in each county is extracted from the ACS dataset.[40] This study gathered data regarding the percentage of individuals without health insurance in each county from resources provided by NIH [41] Data on the number of COVID-19 cases used to calculate the COVID-19 rate was sourced from USAFacts [42].

*3.2 Variables*

Upon obtaining the mobile location data set for 2019 and 2020, the dataset was subsequently filtered to isolate commute trips that are home-based work (HBW) trips. The DMV region was chosen as the case study for this research. The study was conducted at the county level, with all data variables acquired and aggregated accordingly. The DMV region comprises 158 counties, distributed between the District of Columbia, Maryland, and Virginia. Table 1 Offers an overview of the descriptive statistical data of the variables employed in this study.

Table 1 - Descriptive Statistics of Candidate Variables Over The 158 Counties of The Study Area

| Variables (In Each County) | Min | Max | Mean | Median | SD |
| --- | --- | --- | --- | --- | --- |
| Scientific Jobs % | 2.94 | 30.00 | 10.74 | 9.80 | 4.73 |
| Agriculture Jobs % | 0.00 | 12.04 | 1.98 | 1.40 | 2.03 |
| Construction Jobs % | 0.90 | 16.20 | 7.47 | 7.20 | 2.85 |
| Manufacturing Jobs % | 1.27 | 23.02 | 9.83 | 8.75 | 5.43 |
| Whole sale trade Jobs % | 0.00 | 5.40 | 2.00 | 1.88 | 0.92 |
| Retail Trade Jobs % | 4.09 | 20.29 | 11.45 | 11.46 | 2.60 |
| Information Jobs % | 0.00 | 6.52 | 1.56 | 1.52 | 0.85 |
| Food Services Jobs % | 3.11 | 27.77 | 8.52 | 7.80 | 3.39 |
| Public administration Jobs % | 2.13 | 22.38 | 7.69 | 6.96 | 3.71 |
| Median Income (thousands $) | 27.06 | 142.30 | 63.47 | 57.45 | 23.01 |
| Active Mode of Travel % | 0.85 | 23.09 | 6.75 | 6.12 | 3.05 |
| Public Mode of Travel % | 0.00 | 17.82 | 0.67 | 0.02 | 2.23 |
| People Vote Democrat% | 0.21 | 0.91 | 0.46 | 0.44 | 0.14 |
| COVID-19 Rate | 0.00 | 1.60 | 0.17 | 0.10 | 0.21 |
| People with No Health Insurance % | 5.90 | 41.50 | 17.82 | 17.05 | 5.44 |

The independent variable in this study is the rate of trip change, calculated using the following formula:

$$TC = \frac{(T_b - T_a) * 100}{T_b} \quad (1)$$

Where $TC$ represents the change in commuting trips before and after the onset of COVID-19, $T_b$ shows the number of commute trips in March-April, and November of 2019, which is referred to as before COVID-19, where there were no signs of COVID-19, and $T_a$ represents the number of commute trips in March-April, and November of 2020, which is referred to as after COVID-19, where there was peak of trip change in 2020. After calculating the trip change, a positive sign indicates a decrease in the number of trips, while a negative sign shows an increase in the number of trips. A more remarkable positive trip change implies a more significant decrease in the number of trips.

Travel behavior change started in March 2020; before COVID-19 led to widespread closures and disruptions on a global scale, there were not considerable trip changes between months. The peak of the commute trip change between 2019 and 2020 is 121.9% for April. Also, the second highest commute trip change between 2019 and 2020 is related to the last of the year during October and November. In those months, there was a resurgence of COVID-19 Cases [43]. These months are prior to the widespread vaccination rollout across the US. Therefore, it becomes crucial to observe how travel behavior might have altered immediately after the onset of COVID-19 or a comparable catastrophe, before returning to a state of normalcy. At the beginning of the pandemic, the circumstances were unprecedented and uncertain, leading to significant changes in people's behaviors. As time passed, individuals became more acquainted with the situation, resulting in the emergence of new patterns and adaptations. Figure 1 depicts the trajectory of changes in commuting trips throughout 2019 and 2020 in DMV area.

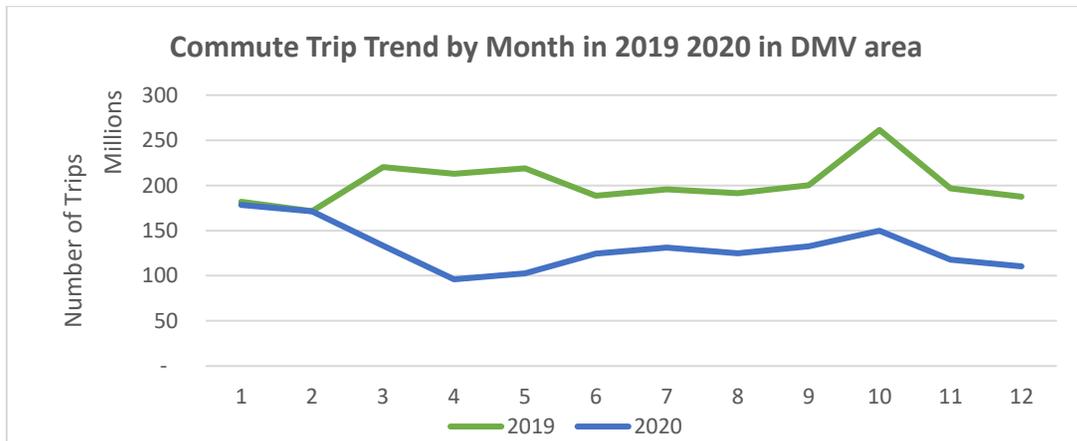

Figure 1 - Commute trip trend by month for 2019 and 2020 in DMV area

In this study, the baseline model examines trip changes before and after the onset of COVID-19. The period before and after COVID-19 is selected as March 15th to April 15th, 2019 and 2020. This selection is based on Figure 2**Error! Reference source not found.**, which illustrates the social distancing index for all U.S. states. As depicted in the figure, there are five stages [44]. Until mid-March, the social distancing index remained relatively stable, indicating minimal changes in behavior during the pre-pandemic period. However, people began altering their routines in the latter half of March. The highest social distancing index values and government-mandated stay-at-home orders occurred between mid-March and mid-April. Also, Percentage of decreased trips from March-April 2019 to 2020 is 107.35, which is among the highest.

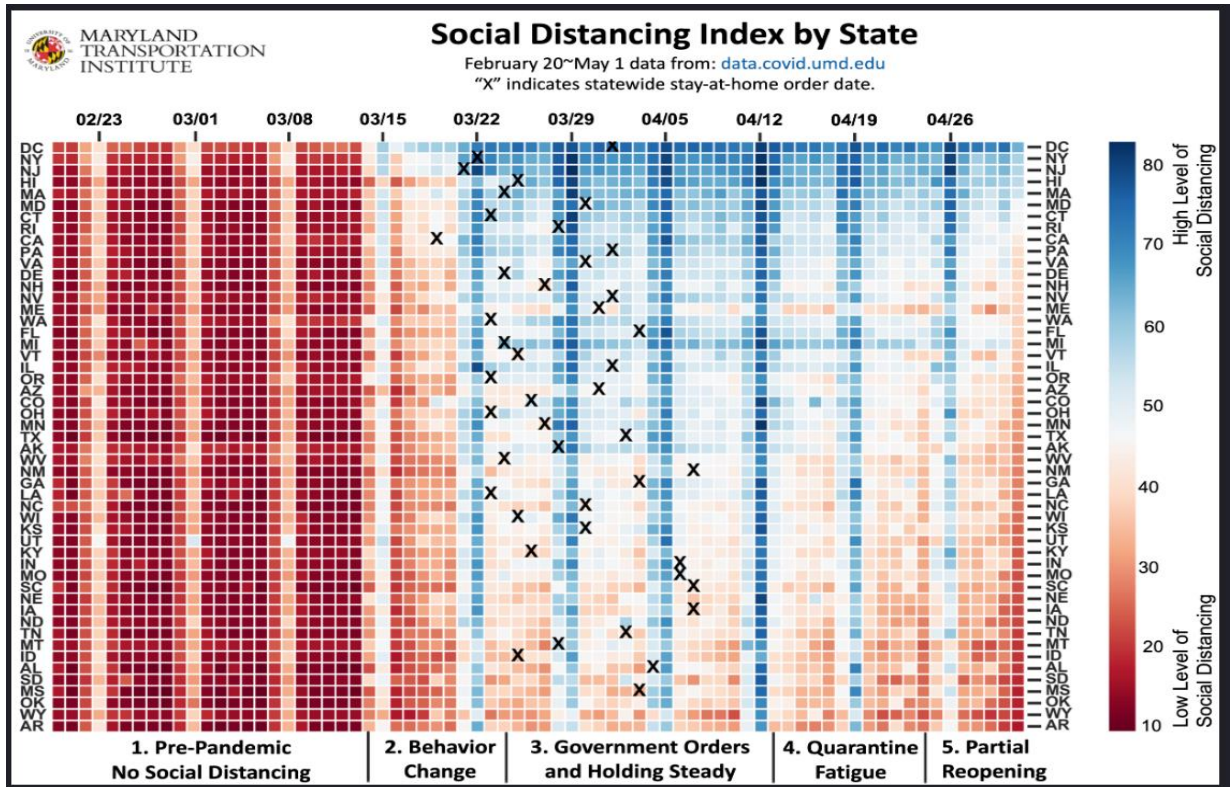

Figure 2 - Social distance index by state [44]

Another time frame examined in this research involves the trip changes between November 2019 and 2020.

*3.3 Methodological Framework*

The methodological framework proposed for this study encompasses two primary sections: explanatory and spatial-analysis models. The OLS regression is used to examine the relationship between multiple explanatory variables and a dependent variable. The equation in this study is given by:

$$Y = \beta_0 + \beta_i * X_i + \varepsilon \qquad (2)$$

where $Y$ is the value for trip change. $X$ represents the Independent variable. $\beta_0$ is the intercept. $\beta_i$ refers related coefficients, and $\varepsilon$ is the random error term. In the explanatory model section, the initial approach entails utilizing OLS models to analyze 15 variables, focusing on the signs of their coefficients and their statistical significance. In this segment, each variable's significance and magnitude across the scenarios' coefficients are meticulously examined and compared. To further refine the model, tests and checks are employed to ascertain the autocorrelation of the model's residuals. Finally, a GWR and MGWR models are implemented as an extended and more advanced version of the standard OLS regression. The GWR and MGWR model facilitates the estimation of local parameters, thereby providing a more nuanced alternative to global parameter estimation [45] [28]. GWR equation is given by:

$$Y_i = \beta_{i0} + \sum_{j=1}^{n} \beta_{ij} * X_{ij} + \varepsilon_i \tag{3}$$

where at areas $i$, $Y_i$ is the value for the trip change. $\beta_{i0}$ is the intercept for areas $i$. $\beta_{ij}$ refers the coefficients of different explanatory variables in different areas, and $X_{ij}$ is the related explanatory variable. $\varepsilon_i$ is the random error. MGWR equation is given by:

$$Y_i = \beta_0(u_i, v_i) + \sum_{j=1}^{m} \beta_{bwj} * (u_i, v_i) X_{ij} + \varepsilon_i \tag{4}$$

The intercept $\beta_0$, independent variable ($X_{ij}$) in the coordinates of each observation ($u_i, v_i$), error term $\varepsilon_i$ and bandwidth $\beta_{bwj}$ for calibrating the $j_{th}$ conditional relationship is defined in the formula.

## 4. Results

### 4.1 Explanatory Model

Table 2 presents the results of OLS models. All variables that appear to impact the dependent variable significantly and do not have multicollinearity are included. Table 2 also displays the VIF values of the variables used in both short-term and long-term model, and there is no indication of multicollinearity. Multicollinearity can vary depending on the context and the specific field of study. In this context, those two variables with a VIF exceeding 5 prove insightful and valuable for interpreting the results. The outcomes associated with the simultaneous inclusion of both variables suggest enhanced performance compared to the use of either one in isolation. Moreover, according to the literature review, a widely accepted threshold for identifying high multicollinearity is a VIF value surpassing 7.5 or 10. [24] [46] [47] The results indicate a positive correlation between the number of people utilizing active transit (walk and bike) for work-related travel and the decrease in commuting trips during the initial phase of COVID-19. Conversely, for the long-term effect of COVID-19 in November, we can see the impact of using active mode on trip change is not significant anymore and instead results showed people who were using bus and rail (public mode) for commuting decreased their trips more than others.

As anticipated, the level of median income exhibited significance and demonstrated a positive correlation with changes in trips. Individuals with higher incomes tended to stay at home more during the early stages of COVID-19 and continued to do so after some time. Conversely, sectors that required continued outdoor operations during the COVID-19 pandemic, such as manufacturing, wholesale trade, and food services, exhibited significant negative relation with trip change, as anticipated [26].

Table 2 - Linear Regression Results for March-April and November

| Dependent Variable | Trip change before and after COVID-19 Short term – (March-April) | | | | Trip change before and after COVID-19 Long term – (November) | | | |
|---|---|---|---|---|---|---|---|---|
| | Est Coeff | T-value | P-Value | VIF | Est Coeff | T-value | P-Value | VIF |
| Intercept | 1.039 | 9.39 | 0.000*** | - | 1.19 | 9.52 | 0.000*** | - |
| Scientific | 0.080 | 0.87 | 0.384 | 5.47 | 0.02 | 0.24 | 0.813 | 5.48 |
| Agriculture | -0.110 | -1.95 | 0.053 | 1.93 | -0.29 | -4.62 | 0.000*** | 1.84 |
| Construction | -0.088 | -1.78 | 0.077 | 1.81 | -0.17 | -2.91 | 0.004** | 1.85 |
| Manufacturing | -0.207 | -4.52 | 0.000*** | 2.80 | -0.17 | -3.28 | 0.001** | 2.81 |
| Wholesale trade | -0.111 | -2.35 | 0.02* | 1.40 | -0.21 | -3.97 | 0.000*** | 1.40 |
| Retail trade | 0.003 | 0.06 | 0.954 | 1.88 | 0.06 | 0.94 | 0.351 | 1.89 |
| Information | -0.219 | -3.20 | 0.002** | 1.68 | -0.41 | -5.24 | 0.000*** | 1.71 |
| Food services | -0.354 | -5.26 | 0.000*** | 1.82 | -0.18 | -2.29 | 0.024* | 1.90 |
| Public administration | -0.135 | -2.37 | 0.019* | 2.31 | -0.22 | -3.35 | 0.001** | 2.30 |
| Median Income | 0.276 | 3.22 | 0.002** | 6.29 | 0.30 | 3.19 | 0.002** | 5.89 |
| Active mode | 0.172 | 2.01 | 0.046* | 2.94 | 0.00 | -0.02 | 0.986 | 2.88 |
| Public mode | -0.054 | -0.58 | 0.564 | 2.87 | 0.25 | 2.46 | 0.015* | 2.41 |
| Democrat voters | -0.015 | -0.29 | 0.771 | 2.18 | -0.12 | -1.96 | 0.050* | 2.34 |
| COVID-19 rate | -0.062 | -1.38 | 0.17 | 1.05 | -0.05 | -1.01 | 0.313 | 1.05 |
| Health uninsured | -0.079 | -1.27 | 0.205 | 1.94 | -0.06 | -0.86 | 0.391 | 1.94 |
| r-squared | 0.551 | | | | 0.524 | | | |
| adj. r-squared: | 0.503 | | | | 0.474 | | | |
| method: | Least Squares | | | | Least Squares | | | |
| No. Observations: | 158 | | | | 158 | | | |
| Df Residuals: | 142 | | | | 142 | | | |
| Df Model: | 15 | | | | 15 | | | |
| AIC: | -313.2 | | | | -269.7 | | | |
| BIC: | -264.2 | | | | -220.7 | | | |

Another variable that got significant sometime after COVID-19 happened is the percentage of people who voted Democrat in a county. Research by Brodeur et al. shows that counties with a higher percentage of self-identified Democrats experienced decreased mobility during the COVID-19 pandemic, affecting all types of trips. [48] However, in the DMV area, which has a significant proportion of government jobs, people went out to work more after six months of the pandemic, possibly influenced by policies encouraging a return to work for government employees. [49] [50] [51] The presence of government jobs in the DMV area aligns with the higher percentage of Democratic voters. The percentage of people who vote democrat in DC, Maryland and Virginia is 92.15%, 65.36%, 54.11%. Moreover, in the District of Columbia, about 25.1% of jobs were in the government sector. In Maryland, government jobs accounted for approximately 19.8% of total employment. In Virginia, government employment constituted around 16.6% of all jobs, where the average percentage is 15.2%. [52] This hypothesis gains further credibility when the same model is applied to the same time frame (November 2019-2020) but in the context of Florida. This association is not observed in Florida, where the percentage of Democratic voters resulted in a decrease in trips. Consequently, in DMV area counties with more Democratic votes demonstrated a higher tendency for individuals to venture out for work during the pandemic. Thus, the coefficient of this variable aligns logically with the observations.

*4.2 Spatial Analysis:*

As the research progresses, the investigation aims to discern whether local effects, such as clustering and spatial autocorrelation, were present from the pandemic's onset or whether these patterns emerged over time. Utilizing GWR and MGWR is essential in this investigation, allowing for the analysis of local rather than global parameters. By deploying the GWR & MGWR model, this study can aptly account for spatial non-stationarity that may otherwise be overlooked by simple global fitting methods, thereby providing a more nuanced understanding of the spatial variance in COVID-19's impact on commuting patterns.

For the following steps, the Presence of residual spatial autocorrelation was inspected using global Moran's I. The spatial analysis of the residuals revealed significant spatial autocorrelation in the model for November. This observation is supported by a Moran's I test ($I= -0.006$, $p= 0.007$), confirming the presence of spatial autocorrelation of residuals. Conversely, the Moran's I test for March to April yielded a p-value of 0.21, which is insufficient to establish spatial autocorrelation in the dataset for this time period.

While describing non-stationary spatial relationships, the GWR mostly confirmed the OLS model's results. The GWR coefficients indicated the presence of spatial variation, as expected. Consequently, these findings suggest that the influence of COVID-19 on commuting behaviors initially displayed global effects. However, over time, as the pandemic progressed, these effects began exhibiting local correlations, hinting at a more complex, geographically dependent impact pattern. The following section will examine the results of the GWR and MGWR model and interpret the spatial distribution for the November case study.

Due to the high sensitivity of the GWR model to multicollinearity, any variables causing such collinearity have been excluded from the model. Within the geographically weighted framework, collinearity may be present in local subsets [53]. The GWR and MGWR analyses were combined with the local collinearity diagnostic tests using the MGWR Python package [54]. Figure 3 shows the exact location of DMV area in the map. The variability of local condition numbers (CN) for the GWR and MGWR models is depicted in Figure 4. The impact of collinearity was significant in certain areas of the GWR model, particularly in the Southern Midwest, where regions exhibited a CN greater than 30. These values indicate a collinear relationship among the predictor variables [55]. The collinearity observed could be attributed to the fixed GWR bandwidth, as it tends to increase collinearity between variables. [56] All of the local MGWR models were revealed to have CNs of less than 24 which is acceptable.

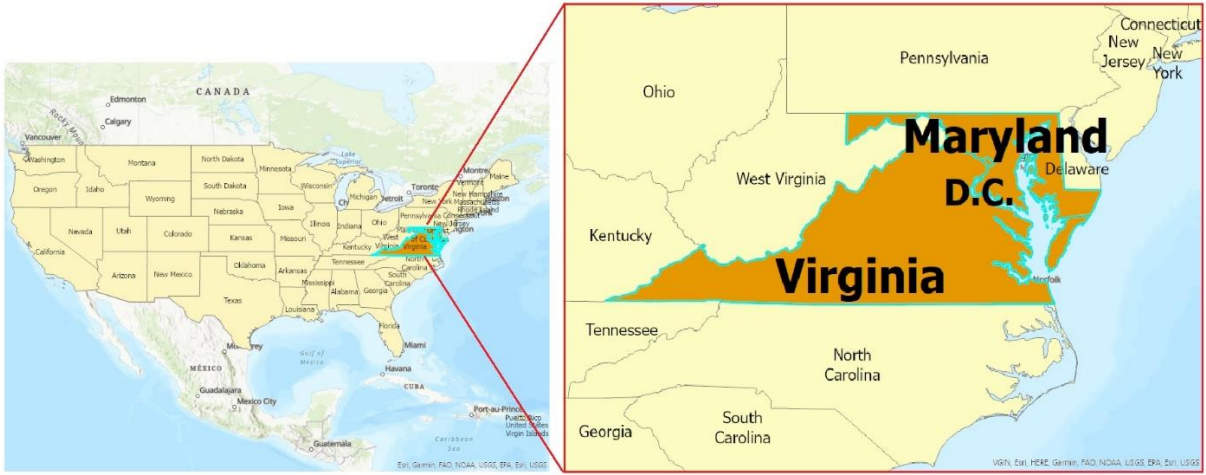

Figure 3 - Exact location of case study (DMV area)

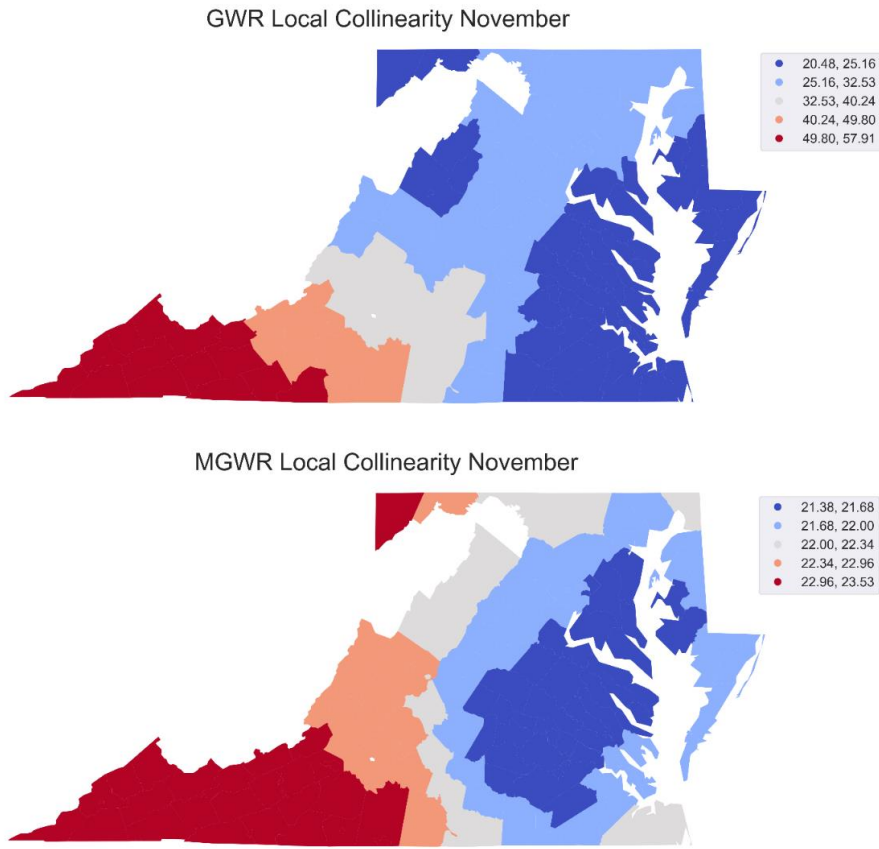

Figure 4 - Diagnostic tests of the local collinearity for the GWR & MGWR

Trip Change varies across the DMV area, with the central southern counties of Maryland and the District of Columbia having the highest trip change (Figure 5). Getis-Ord Gi* Hot Spot analysis shows high clustering of trip change in the middle southern counties of Maryland and the District of Columbia, while there are cold (low) clusters of trip change in some counties and regions of Virginia (Figure 6). The analysis reveals interesting geographical distributions across the DMV area. It is found that the percentage of individuals with agriculture jobs and COVID-19 rate is particularly high in the southwest of Virginia. Conversely, regions with a high concentration of public administration Jobs are predominantly found in the District of Columbia and southern Maryland. These areas also report a higher population of individuals without insurance and those who depend on public modes of transportation. These areas coincide with regions where the trip change was most pronounced. However, the distribution of food services and wholesale trade Jobs across the DMV area appears to be more evenly dispersed.

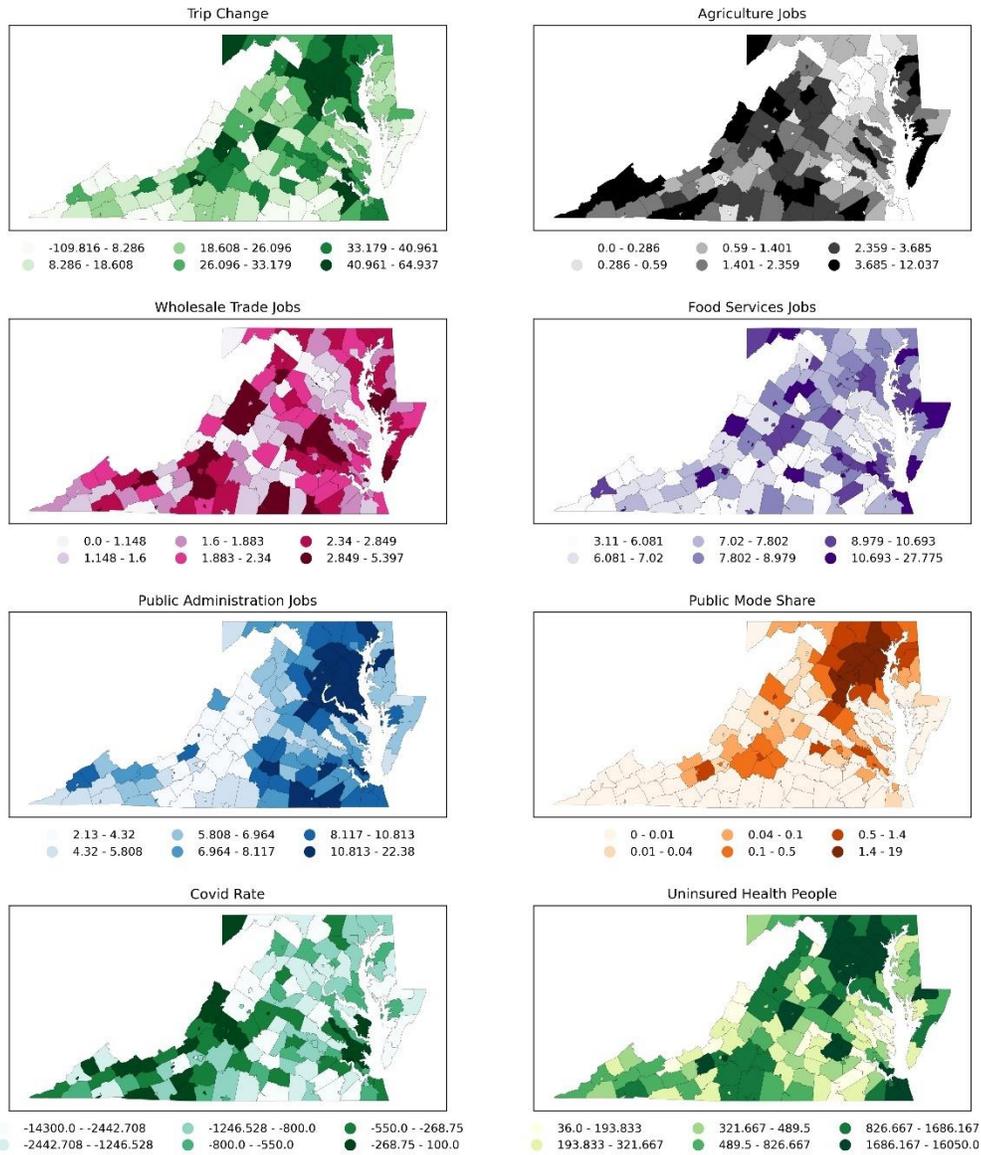

Figure 5 - Spatial Distribution of Variables

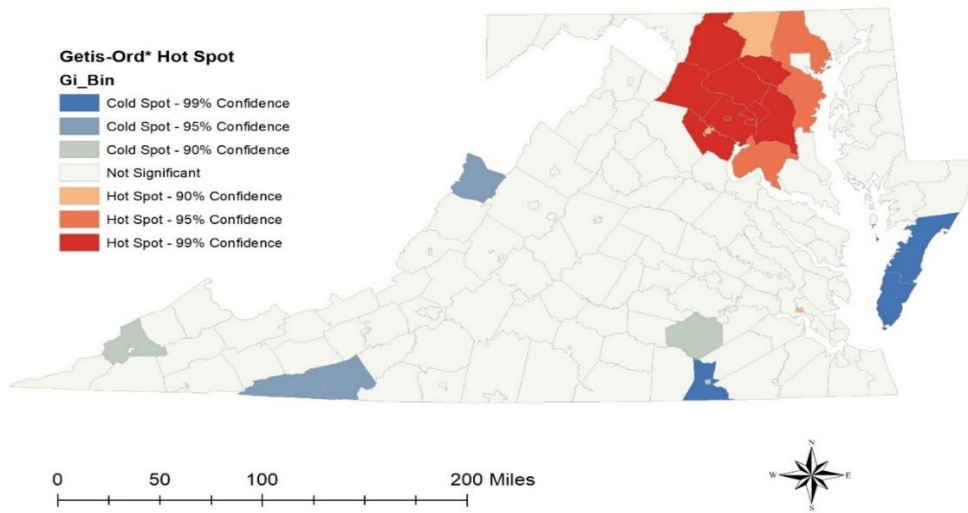

Figure 6 - Geographical Clusters of Counties from Getis-Ord Gi* Statistics of Trip Change

Figure 7 and Figure 8 show the intercept's GWR & MGWR coefficient estimates and each covariate. Coefficient surface comparisons can help to improve understanding of spatial and scale variations. A positive coefficient for an independent variable means that the dependent and independent variables are positively related in that location: as the independent variable increases, the dependent variable also increases. A negative coefficient means the variables are negatively related: as the independent variable increases, the dependent variable decreases. The coefficient maps also show how these relationships vary in space. In the GWR model, the coefficient estimates show a negative relationship between whole sale trade Jobs and Trip Change in the east and partially in the middle but a positive in the west side. Conversely, the coefficient estimates show a positive relationship between public administration Jobs and Trip Change in the north and partially in the east but a negative relationship between trip change and public administration Jobs in the west side. Similarly, the MGWR model confirmed the link between trip change and public administration and whole sale trade jobs. The coefficient estimates of the GWR model for the variable uninsured health people have a positive relationship with the dependent variable with the variations in estimates in the middle and southeast of the DMV area. The MGWR estimate of the uninsured health people on the west side confirms this. The GWR and MGWR model coefficient estimates show a negative relationship between "agriculture jobs" and "public mode share" and trip change.

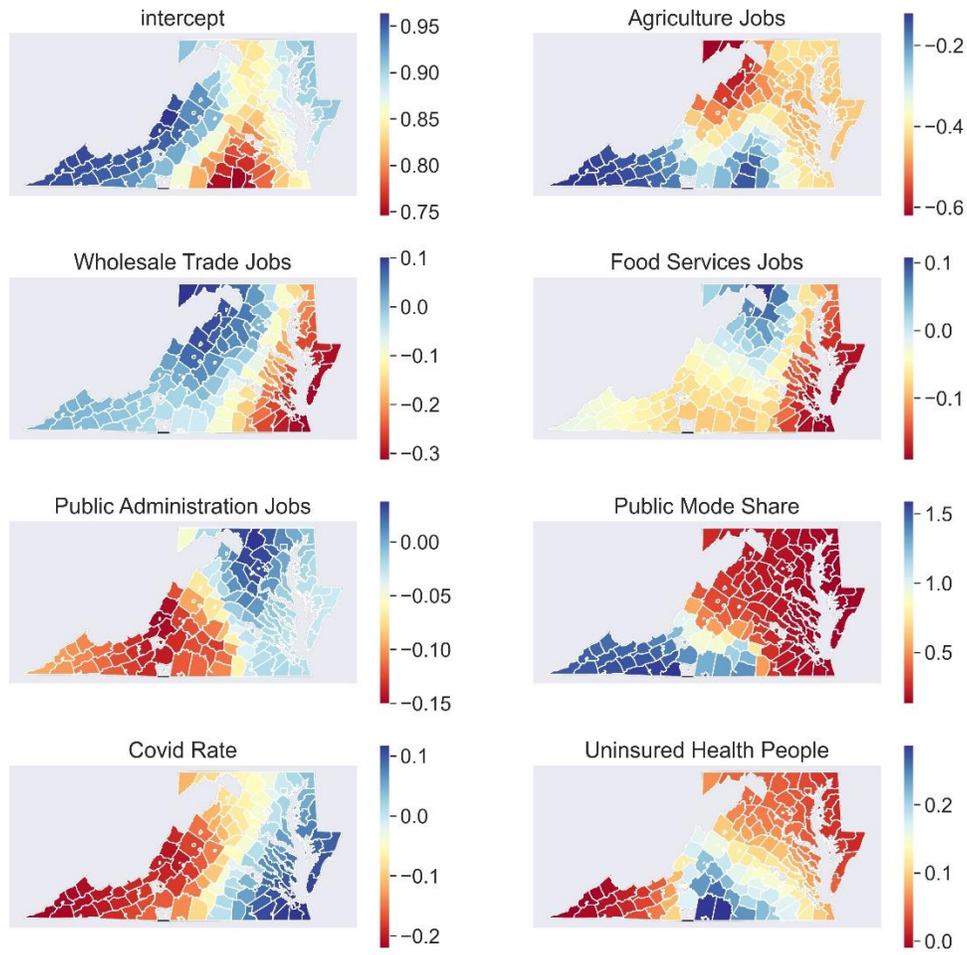

Figure 7 - Spatial Distribution of Coefficients of GWR

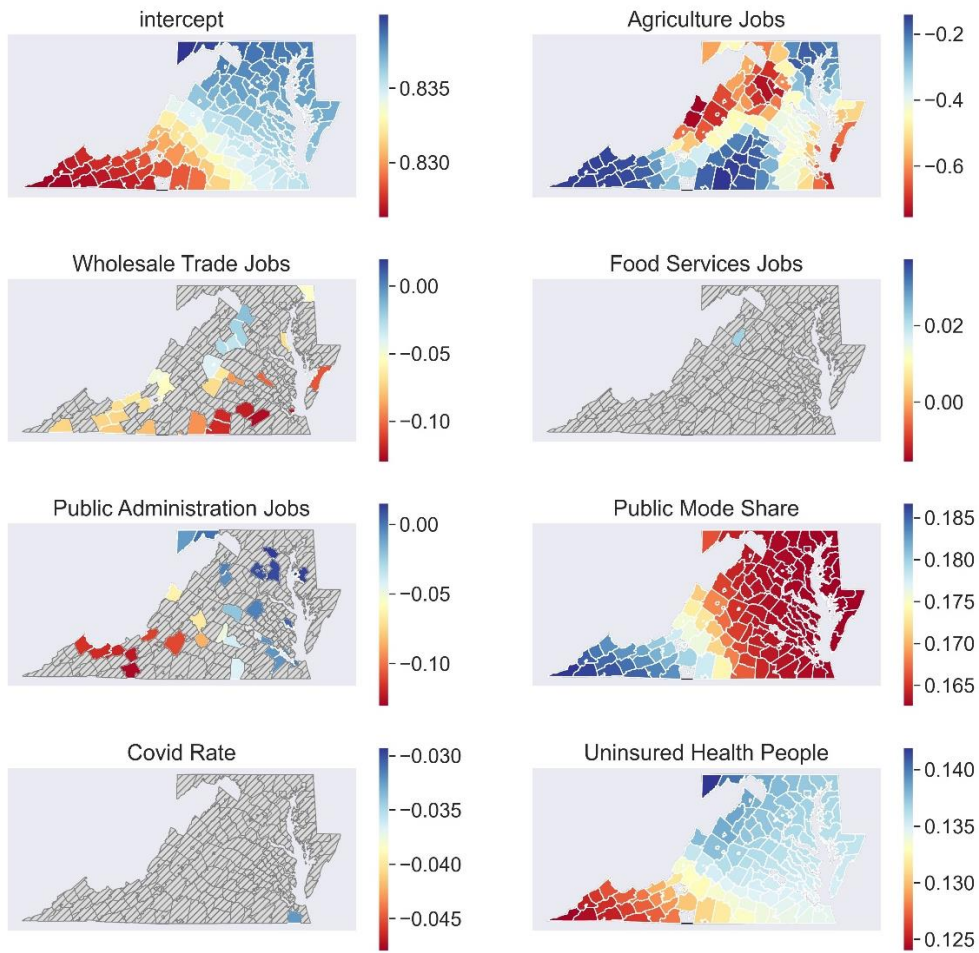

Figure 8 - Spatial Distribution of Coefficients of MGWR

The findings (Table 3) show that "Agriculture Jobs" has a strong significant relationship with trip change Figure 8 with a median coefficient value of (-0.42) and operates at a highly localized scale of 44. However, the MGWR bandwidth of 156 shows the global variation of the "Public Mode Share" and "Uninsured Health People" variable compared to the GWR bandwidth of 112. The relationship between these two variables and trip change is strongly significant. The relationship between Wholesale Trade and trip change is significant in mid-south and west (with negative associations)

Table 3 - Summary Statistics for MGWR Parameter Estimates

| Dependent Variable | Trip change before and after COVID-19 Long term – (November) | | | | |
|---|---|---|---|---|---|
| | Bandwidth | STD | Min | Median | Max |
| Intercept | 156 | 0.00 | 0.83 | 0.84 | 0.84 |
| Agriculture | 44 | 0.18 | -0.76 | -0.42 | -0.14 |
| Wholesale trade | 107 | 0.04 | -0.13 | -0.07 | 0.02 |
| Food services | 151 | 0.01 | -0.02 | 0.01 | 0.04 |
| Administration | 126 | 0.05 | -0.13 | -0.02 | 0.02 |
| Public mode | 156 | 0.01 | 0.16 | 0.16 | 0.19 |
| COVID-19 rate | 156 | 0.01 | -0.05 | -0.03 | -0.03 |
| Health uninsured | 156 | 0.00 | 0.12 | 0.14 | 0.14 |
| r-squared | 0.504 | | | | |
| adj. r-squared: | 0.424 | | | | |
| AIC: | -249.3 | | | | |
| BIC: | -241.2 | | | | |

## 5. Discussion

COVID-19 was a novel and unexpected challenge for which the world was unprepared. This pandemic has had broad effects on people's lives in various aspects, including economic and social dimensions. Transportation and travel behavior were also significantly impacted by these changes. One of these impacts was on people's decisions regarding both commute and non-commute trips in the short and long term. There is no guarantee that something like this pandemic will not happen again in the world. This uncertainty could lead to fundamental changes, such as an increased trend of people working from home. Because of this, our study is crucial to understand the factors influencing travel behavior and how they operate.

Furthermore, examining the Temporal and Spatial Varying Impacts on Commute Trip Change in this study can enhance our understanding of factors not directly accounted for in ordinary models. This exploration aids in identifying additional variables that may significantly influence changes in travel behavior, particularly in relation to the spatial aspects of the area. Analyzing the relationship between variables at a geographical level using spatial models reveals intriguing findings, providing insights into how different areas affect each other. These findings can be applied in various scenarios, such as examining the relationships between zip codes, counties, or even states and their neighbors, elucidating the ways in which they impact one another.

The approach employed in this study to analyze travel behavior based on geographical features proves insightful. As demonstrated, travel behavior exhibits a geographical orientation, emphasizing the importance of considering geographic constraints in other travel behavior studies. Employing innovative local regression models such as GWR and MGWR in this study to assess the spatial and temporal impacts of COVID-19 on commute trips, utilizing extensive mobile location data, can illuminate previously overlooked factors, contributing valuable insights not captured in prior research.

This study's limitation is the absence of a corresponding table to specify the job description for each commute trip, which may require travel surveys or utilizing POI approaches. However, having this information at larger study scales, such as national or state, would be challenging. Additionally, examining travel behavior and job descriptions focused at a more detailed level, such as one town or single county scale, may provide more accurate insights due to local variations. The heterogeneity of travel behavior and job description in different regions is not considered in this study; despite the geographical closeness of the DMV area, their unique urban networks influence movement and employment patterns. The densely populated urban core of DC differs from the more suburban and rural areas of Maryland and Virginia. However, it is crucial to acknowledge that this may give a less comprehensive overview of broader regional dynamics compared to state-level analysis.

From two perspectives, this study is important for future research. Firstly, in emergency situations like COVID-19, this study can assist researchers and policymakers in gaining a more comprehensive insight into the situation. Secondly, the analysis of impactful parameters on teleworking and remote work, which can be undertaken in the next phase of this paper, adds another dimension to its significance.

Moreover, for future research, a national-level study analyzing commute trip changes could offer valuable insights into how COVID-19 may have impacted states with varying regulations and cultures. For instance, it was observed that a higher percentage of Democratic voters in the DMV area had a negative relation, but a positive relationship with trip change in Florida. Conducting more comprehensive studies exploring the long-term effects of post-COVID-19 would also be beneficial. For example, examining how COVID-19 impacted commute trips two years after its onset, when the pandemic has largely subsided and life has mostly returned to normal. It would be interesting to see how COVID-19 might have altered travel behavior, and how people react even when the pandemic has almost come to an end.

## 6. Summary and Conclusion

This study examines the temporal and spatial varying impacts on commute trip change due to COVID-19 in the DMV area. Based on the literature review, the key research gap is identified, and a methodological framework is proposed to apply regression models to understand the influential variables on trip change.

Firstly, mobile device location data was collected for March, April, and November of 2019 and 2020, and then filtered for Home-based Work trips in the DMV area. Job sector, sociodemographic, and econometric variables for each county were incorporated into the dataset. The final dataset contains all the information for variables and the trip change from 2019 to 2020 for each county.

During the initial days of COVID-19, in response to the stay-at-home orders, the use of active modes of transportation became a significant factor in reducing commute trips. However, after six months, this reduction shifted to people who were using public transportation. It becomes crucial to understand how to manage different modes of transport for individuals during such disasters.

Additionally, counties with a higher percentage of jobs requiring field labor, such as manufacturing, wholesale trade, and food services, exhibited a negative correlation with trip change, suggesting a reduction in trip changes due to COVID-19. Conversely, the higher-income jobs have a positive relationship with trip change. As an equity concern, it is crucial for governors and policymakers to ensure a situation where safety from a disaster is accessible not only to people with financial means but to everyone.

Surprisingly, the results indicate that counties with a higher percentage of Democratic voters did not reduce their commute trips as others, even though Democratic voters generally complied with the stay-at-home order and decreased their daily trips. This discrepancy is related to their job sector, which is more government-focused.

The analysis of spatial correlation over time of commute trip changes reveals that initially, COVID-19's influence on commuting behaviors exhibited global effects. However, as the pandemic evolved, these effects started to show local correlations, suggesting a more intricate pattern of impact that depended on geographic factors. The GWR and MGWR results show how agriculture and wholesale trade jobs operate at a highly localized level and vary across the DMV area. Public administration jobs have a positive relation with trip change in the DC area and a negative relation in Virginia. Moreover, the percentage of uninsured people exhibits a higher negative relation with trip change in the DC area.

**ACKNOWLEDGMENTS**
The authors would like to thank and acknowledge their partners and data sources in this effort: (1) parts of the algorithms developed and validated using partial financial support from the U.S. Department of Transportation Federal Highway Administration (2) mobile device location data provider partners

**AUTHOR CONTRIBUTIONS**
The authors confirm their contribution to the paper as follows: study conception and design: S Saleh Namadi, B. Tahmasbi, A. Mehditabrizi, D. Niemeier; data collection and preparation: S. Saleh Namadi, A. Darzi; analysis and interpretation of results: S Saleh Namadi, B. Tahmasbi, A. Mehditabrizi; draft manuscript preparation: S Saleh Namadi, A. Mehditabrizi, B. Tahmasbi, A. Darzi. All authors reviewed the results and approved the final version of the manuscript.

The authors declare that they have no conflict of interest.

**References**

[1]     J. Pöschl, "The effects of the corona shock on the banking sector and the real economy," Economic Memo, 2020.

[2]     M. Bognanni, D. Hanley, D. Kolliner, and K. Mitman, "Economic activity and covid-19 transmission: Evidence from an estimated economic-epidemiological model," *Finance and Economics Discussion Series*, vol. 91, 2020.

[3]     Q. Li *et al.*, "Early Transmission Dynamics in Wuhan, China, of Novel Coronavirus–Infected Pneumonia," *N Engl J Med*, vol. 382, no. 13, pp. 1199–1207, Mar. 2020, doi: 10.1056/NEJMoa2001316.

[4]     J. Csse, "Covid-19 data repository by the center for systems science and engineering (csse) at johns hopkins university," *CSSE, Editor*, 2020.

[5]     M. Chinazzi *et al.*, "The effect of travel restrictions on the spread of the 2019 novel coronavirus (COVID-19) outbreak," *Science*, vol. 368, no. 6489, pp. 395–400, 2020.

[6]     M. U. Kraemer *et al.*, "The effect of human mobility and control measures on the COVID-19 epidemic in China," *Science*, vol. 368, no. 6490, pp. 493–497, 2020.

[7]     K. Gkiotsalitis and O. Cats, "Public transport planning adaption under the COVID-19 pandemic crisis: literature review of research needs and directions," *Transport Reviews*, vol. 41, no. 3, pp. 374–392, 2021.

[8]     P. R. Stopher and S. P. Greaves, "Household travel surveys: Where are we going?," *Transportation Research Part A: Policy and Practice*, vol. 41, no. 5, pp. 367–381, 2007.

[9] N. Schuessler and K. W. Axhausen, "Processing raw data from global positioning systems without additional information," *Transportation Research Record*, vol. 2105, no. 1, pp. 28–36, 2009.

[10] D. M. Harrington and M. Hadjiconstantinou, "Changes in commuting behaviours in response to the COVID-19 pandemic in the UK," *Journal of transport & health*, vol. 24, p. 101313, 2022.

[11] L. Ecke, M. Magdolen, B. Chlond, and P. Vortisch, "How the COVID-19 pandemic changes daily commuting routines–Insights from the German Mobility Panel," *Case Studies on Transport Policy*, vol. 10, no. 4, pp. 2175–2182, 2022.

[12] A. Bick, A. Blandin, and K. Mertens, *Work from home after the COVID-19 outbreak*. Federal Reserve Bank of Dallas, Research Department, 2020.

[13] M.-J. O. Kalter, K. T. Geurs, and L. Wismans, "Post COVID-19 teleworking and car use intentions. Evidence from large scale GPS-tracking and survey data in the Netherlands," *Transportation Research Interdisciplinary Perspectives*, vol. 12, p. 100498, 2021.

[14] N. Chankaew, A. Sumalee, S. Treerapot, T. Threepak, H. W. Ho, and W. H. Lam, "Freight traffic analytics from national truck GPS data in Thailand," *Transportation research procedia*, vol. 34, pp. 123–130, 2018.

[15] S. Çolak, L. P. Alexander, B. G. Alvim, S. R. Mehndiratta, and M. C. González, "Analyzing Cell Phone Location Data for Urban Travel: Current Methods, Limitations, and Opportunities," *Transportation Research Record*, vol. 2526, no. 1, pp. 126–135, Jan. 2015, doi: 10.3141/2526-14.

[16] R. Ahas, S. Silm, E. Saluveer, and O. Järv, "Modelling home and work locations of populations using passive mobile positioning data," *Location Based Services and TeleCartography II: From sensor fusion to context models*, pp. 301–315, 2009.

[17] N. Sari Aslam, T. Cheng, and J. Cheshire, "A high-precision heuristic model to detect home and work locations from smart card data," *Geo-spatial Information Science*, vol. 22, no. 1, pp. 1–11, 2019.

[18] X. Yang, Z. Fang, L. Yin, J. Li, Y. Zhou, and S. Lu, "Understanding the spatial structure of urban commuting using mobile phone location data: a case study of Shenzhen, China," *Sustainability*, vol. 10, no. 5, p. 1435, 2018.

[19] Y. Xu, S.-L. Shaw, Z. Zhao, L. Yin, Z. Fang, and Q. Li, "Understanding aggregate human mobility patterns using passive mobile phone location data: A home-based approach," *Transportation*, vol. 42, pp. 625–646, 2015.

[20] K. S. Kung, K. Greco, S. Sobolevsky, and C. Ratti, "Exploring universal patterns in human home-work commuting from mobile phone data," *PloS one*, vol. 9, no. 6, p. e96180, 2014.

[21] S. Engle, J. Stromme, and A. Zhou, "Staying at home: mobility effects of covid-19," *Available at SSRN 3565703*, 2020.

[22] X. Huang, J. Lu, S. Gao, S. Wang, Z. Liu, and H. Wei, "Staying at home is a privilege: Evidence from fine-grained mobile phone location data in the United States during the COVID-19 pandemic," *Annals of the American Association of Geographers*, vol. 112, no. 1, pp. 286–305, 2022.

[23] L. Zhang *et al.*, "Interactive COVID-19 mobility impact and social distancing analysis platform," *Transportation Research Record*, vol. 2677, no. 4, pp. 168–180, 2023.

[24] J. Jiao, M. Bhat, and A. Azimian, "Measuring travel behavior in Houston, Texas with mobility data during the 2020 COVID-19 outbreak," *Transportation letters*, vol. 13, no. 5–6, pp. 461–472, 2021.

[25] S. Pourfalatoun and E. E. Miller, "Effects of covid-19 pandemic on use and perception of micro-mobility," *Available at SSRN 4113031*, 2022.

[26] T. Santanam, A. Trasatti, H. Zhang, C. Riley, P. Van Hentenryck, and R. Krishnan, "Changes in Commuter Behavior from COVID-19 Lockdowns in the Atlanta Metropolitan Area," *arXiv preprint arXiv:2302.13512*, 2023.

[27] A. Mollalo, B. Vahedi, and K. M. Rivera, "GIS-based spatial modeling of COVID-19 incidence rate in the continental United States," *Science of the total environment*, vol. 728, p. 138884, 2020.

[28] C. Brunsdon, A. S. Fotheringham, and M. E. Charlton, "Geographically weighted regression: a method for exploring spatial nonstationarity," *Geographical analysis*, vol. 28, no. 4, pp. 281–298, 1996.

[29] P. Goovaerts, "Geostatistical analysis of health data: State-of-the-art and perspectives," in *GeoENV VI–Geostatistics for environmental applications: Proceedings of the sixth European conference on Geostatistics for environmental applications*, Springer, 2008, pp. 3–22.

[30] J. Jiao, Y. Chen, and A. Azimian, "Exploring temporal varying demographic and economic disparities in COVID-19 infections in four US areas: Based on OLS, GWR, and random forest models," *Computational urban science*, vol. 1, pp. 1–16, 2021.

[31] Y. Liu, Z. He, and X. Zhou, "Space-time variation and spatial differentiation of COVID-19 confirmed cases in Hubei Province based on extended GWR," *ISPRS international journal of geo-information*, vol. 9, no. 9, p. 536, 2020.

[32] X. Wu and J. Zhang, "Exploration of spatial-temporal varying impacts on COVID-19 cumulative case in Texas using geographically weighted regression (GWR)," *Environmental Science and Pollution Research*, vol. 28, pp. 43732–43746, 2021.

[33] "NHTS NextGen OD Data." Accessed: May 26, 2023. [Online]. Available: https://nhts.ornl.gov/od/

[34] M. Yang *et al.*, "Big-Data Driven Framework to Estimate Vehicle Volume Based on Mobile Device Location Data," *Transportation Research Record*, p. 03611981231174240, 2023.

[35] Y. Pan *et al.*, "Residency and worker status identification based on mobile device location data," *Transportation Research Part C: Emerging Technologies*, vol. 146, p. 103956, 2023.

[36] A. Kabiri *et al.*, "Elaborated Framework for Duplicate Device Detection from Multisourced Mobile Device Location Data," *Transportation Research Record*, p. 03611981231201114, Nov. 2023, doi: 10.1177/03611981231201114.

[37] H. Gong, C. Chen, E. Bialostozky, and C. T. Lawson, "A GPS/GIS method for travel mode detection in New York City," *Computers, Environment and Urban Systems*, vol. 36, no. 2, pp. 131–139, 2012.

[38] L. Stenneth, O. Wolfson, P. S. Yu, and B. Xu, "Transportation mode detection using mobile phones and GIS information," in *Proceedings of the 19th ACM SIGSPATIAL international conference on advances in geographic information systems*, 2011, pp. 54–63.

[39] M. Yang, Y. Pan, A. Darzi, S. Ghader, C. Xiong, and L. Zhang, "A data-driven travel mode share estimation framework based on mobile device location data," *Transportation*, vol. 49, no. 5, pp. 1339–1383, 2022.

[40] U. C. Bureau, "American community survey (acs)," *The United States Census Bureau nd https://www. census. gov/programs-surveys/acs (accessed May 5, 2021)*, 2016.

[41] "for | HDPulse Data Portal." Accessed: May 26, 2023. [Online]. Available: https://hdpulse.nimhd.nih.gov/data-portal/healthcare

[42] "USAFacts | Nonpartisan Government Data," USAFacts. Accessed: May 26, 2023. [Online]. Available: https://usafacts.org/

[43] "WHO Coronavirus (COVID-19) Dashboard." Accessed: May 26, 2023. [Online]. Available: https://covid19.who.int

[44] M. T. Institute, "University of Maryland COVID-19 impact analysis platform," *University of Maryland, College Park, USA [Internet]*, 2020.

[45] A. S. Fotheringham, C. Brunsdon, and M. Charlton, *Geographically weighted regression: the analysis of spatially varying relationships*. John Wiley & Sons, 2003.

[46] I. Ahmad *et al.*, "Spatial configuration of groundwater potential zones using OLS regression method," *Journal of African Earth Sciences*, vol. 177, p. 104147, 2021.

[47] J. D. Curto and J. C. Pinto, "The corrected vif (cvif)," *Journal of Applied Statistics*, vol. 38, no. 7, pp. 1499–1507, 2011.

[48] A. Brodeur, I. Grigoryeva, and L. Kattan, "Stay-at-home orders, social distancing, and trust," *Journal of Population Economics*, vol. 34, no. 4, pp. 1321–1354, 2021.
26

[49] "M-20-23.pdf." Accessed: May 29, 2023. [Online]. Available: https://www.whitehouse.gov/wp-content/uploads/2020/04/M-20-23.pdf

[50] "Post Hybrid Work Environment Guidance," U.S. Office of Personnel Management. Accessed: May 29, 2023. [Online]. Available: https://www.opm.gov/policy-data-oversight/future-of-the-workforce/post-hybrid-work-environment-guidance/

[51] "Biden calls for 'vast majority' of federal employees to return to office as COVID-19 conditions improve," Federal News Network. Accessed: May 29, 2023. [Online]. Available: https://federalnewsnetwork.com/workforce/2022/03/biden-urges-more-federal-employees-to-return-to-the-office-as-pandemic-conditions-improve/

[52] "Profile of Federal Civilian Non-Seasonal Full-Time Employees," U.S. Office of Personnel Management. Accessed: May 29, 2023. [Online]. Available: https://www.opm.gov/policy-data-oversight/data-analysis-documentation/federal-employment-reports/reports-publications/profile-of-federal-civilian-non-postal-employees/

[53] D. Wheeler and M. Tiefelsdorf, "Multicollinearity and correlation among local regression coefficients in geographically weighted regression," *J Geograph Syst*, vol. 7, no. 2, pp. 161–187, Jun. 2005, doi: 10.1007/s10109-005-0155-6.

[54] T. M. Oshan, Z. Li, W. Kang, L. J. Wolf, and A. S. Fotheringham, "mgwr: A Python implementation of multiscale geographically weighted regression for investigating process spatial heterogeneity and scale," *ISPRS International Journal of Geo-Information*, vol. 8, no. 6, p. 269, 2019.

[55] I. Gollini, B. Lu, M. Charlton, C. Brunsdon, and P. Harris, "GWmodel: an R package for exploring spatial heterogeneity using geographically weighted models," *arXiv preprint arXiv:1306.0413*, 2013.

[56] T. M. Oshan and A. S. Fotheringham, "A comparison of spatially varying regression coefficient estimates using geographically weighted and spatial-filter-based techniques," *Geographical Analysis*, vol. 50, no. 1, pp. 53–75, 2018.